# Analysis of accuracy and efficiency of neural networks to simulate Navier-Stokes fluid flows with obstacles

**Rui Hespanha[1*], Elliot McGuire[2*], João Hespanha[3]**

[1]Dos Pueblos High School, Goleta, California, USA
[2]Dos Pueblos High School, Goleta, California, USA
[3]Department of Electrical Engineering, University of California at Santa Barbara, USA
* These authors contributed equally to this work.

**Student Authors**
Rui Hespanha, Dos Pueblos High School
Elliot McGuire, Dos Pueblos High School

**OVERVIEW:** The authors trained several neural networks on data obtained from conventional physics-based numerical simulations of fluid flows. They then evaluated the networks' performance in predicting fluid flows on environments not in the training data.

**SUMMARY (Abstract)**

Conventional fluid simulations can be time consuming and energy intensive. We researched the viability of a neural network for simulating incompressible fluids in a randomized obstacle-heavy environment, as an alternative to the numerical simulation of the Navier-Stokes equation. We hypothesized that the neural network predictions would have a relatively low error for simulations over a small number of time steps, but errors would eventually accumulate to the point that the output would become very noisy. Over a rich set of obstacle configurations, we achieved a root mean square error of 0.32% on our training dataset and 0.36% on a testing dataset. These errors only grew to 1.45% and 2.34% at $t$ = 10 and, 2.11% and 4.16% at timestep $t$ = 20. We also found that our selected model (center **Figure 1**) was approximately 8,800 times faster at predicting the flow than a conventional simulation. Our findings indicate neural networks can be extremely useful at simulating fluids in obstacle-heavy environments. Useful applications include modeling forest fire smoke, pipe fluid flow, and underwater/flood currents.

**INTRODUCTION**

The Navier-Stokes partial differential equation has long been used to model fluid dynamics. Although accurate, numerical simulations of this equation are computationally intensive and consume a lot of energy. We propose to train a neural network on data collected from Navier-Stokes-based fluid approximate numerical simulations to predict the fluid velocity profile and investigate the accuracy and speed of the neural network predictions of the fluid flow, when compared to a conventional numerical simulation. We consider environments with a large number of obstacles that interact with the fluid, and we specifically want to study the neural network's ability to predict fluid velocities for obstacle configurations that are not in the set of data used to train the neural network.

There has been a growing interest in using neural networks in solving fluid dynamics problems [1]. Preeminent research papers in the field investigate the introduction of physical laws into the training process (an approach commonly known as physics-informed neural networks or PINNs) [2], non-turbulent flow simulations with moving objects [3], integrate novel techniques for training networks [4], and utilize specialized fluid dynamics simulations [5]. We chose to research the efficiency and accuracy of a neural network to predict fluid flows in an obstacle-filled environment, which is a problem that has not been extensively studied using classical neural network training (non-PINN). We were motivated by [6], which states that the classical black-box-style neural networks we use are especially effective in highly constrained environments (such as ones with obstacles), particularly with similar sets of training and testing data (such as our paper)

We hypothesized that our network would be relatively accurate at making predictions over short time horizons, but performance would degrade over time, leading to "noisy" predictions of the speed profile. In reality, our models achieved 99.68-99.64% RMSE accuracy and this accuracy degraded to just 97.89-95.84% after 20 timesteps, with a speed increase 8,800x over the Navier-Stokes fluid simulation in 2 dimensions (see networks 1-5 in **Figure 2**). Rather than generating "noisy" predictions, the neural network actually produced velocity profiles that were overly smooth. This indicates that our method of fluid simulation is reliable and fast for complex environments with incompressible fluid flow and could be expanded upon to provide approximations for turbulent flow patterns or more general physical environments.

**RESULTS**

The neural networks' ability to predict the fluid flow profile was evaluated over a large number of obstacle configurations and, for each configuration, we compared the difference between the fluid velocities predicted by the neural network with those computed by a numerical simulation of the Navier-Stokes equation. The percentage errors reported here refer to the root-mean-square-error (RMSE) averaged over the whole spatial simulation domain, as a fraction of the largest fluid speed over the same domain. We consider multiple neural networks with different numbers of layers and nodes per layer (see x-axis labels in **Figure 2**). For most neural networks, we observed errors in the 2-3% range (see y-axis labels in **Figure 2**). Smaller networks, such as the first three shown in **Figure 2**, computed prediction of the fluid velocities 3,500 to above 8,000 times faster than the conventional simulation. The main limiting factor on network computation speed was width (i.e., number of nodes per layer) despite this parameter actually showing negligible performance improvements. Larger networks (4 & 5 in **Figure 2**) required larger computation times, with both being less than 3,500 times faster, though total network size was not the main factor contributing to speed as the fastest network had a medium size and low width. The difference in error between training and testing data is relatively low: at the end of 19 frame predictions, the network corresponding to the fourth column of **Figure 2** achieves an error of 1.48% error on the training data, which only grows to 2.39% on the testing data. We recall that the testing data set was entirely composed of obstacle configurations that were not present in the training data. Our hypothesis that the predictions would become noisier over time was incorrect. Instead, the neural network predictions for the fluid velocities were overly "smooth" and the error was less than we expected.

**DISCUSSION**

The neural networks generated fluid predictions much faster than numerical simulations with an error that increased over time. Part of this problem could be due to the over-smoothing effect the network has, leading to lower turbulence than in actuality. used in our simulations. It's particularly clear that simulating Predicting fluid dynamics with obstacles over relatively short time horizons can be done very efficiently by a neural network. Over a longer time frame, error growth magnifies, which effectively limits the use of this approach over very long time spans. We were limited in terms of the size of the networks as our GPU was a NVIDIA GeForce 3080 with 10 gigabytes of GPU memory. Another limitation due to low processing power was the low resolution of our simulation. We did not have the processing power to simulate higher resolutions or 3 spatial dimensions in this study. We conjecture that we would encounter larger errors in 3 dimensions due to flow turbulence.

This approach could be extended to modeling fluids that behave qualitatively similar to incompressible fluids under certain conditions including water and smoke (at 1 atmosphere of pressure and earth-surface temperatures) in obstacle-filled environments such as pipes and forests. It also opens the door to cheaper weather modeling that can be performed using trained networks developed through this process (although more research would need to be done). Since our simulation provides a way to cheaply simulate fluid flow around obstacles in short time frames, aerodynamic properties can be calculated by measuring the force exerted from the fluid onto static objects. Similarly, simulating water in a pipe would only require augmenting the spatial dimensional of our simulation and increasing its resolution. These simulations could be utilized to rapidly visualize patterns in fluid flow for given environments.

Future studies are needed to determine the accuracy of this approach in higher resolutions and in 3-dimensional domains. Additionally, object sizes and shapes could be changed to determine how generalizable the approach is. Further investigation is also needed to determine whether this approach can be extended to fluids subject to external forces (such as buoyancy), varying flow parameters, and obstacle shapes. We are interested in specializing this work to specific application domains, such as real-time graphics, weather prediction and modeling, blood flow modeling, and floodwater path modeling. situations. Other applications of our work could be in

To conclude, our simulation is rapid but accumulates error over long periods of time. Our model's unique feature to simulate an environment with obstacle at varying locations allows our model to predict fluid flow patterns in cluttered environments. Our model is useful at modeling any situation with incompressible fluid interacting with an obstacle-filled environment. For air or smoke, modeling fluids in urban environments, forests, or in aerodynamics would work well. For water, modeling its interaction in current, in pipes, or around ships would work well with our simulation.

**METHODS**

To train the neural network, 1,000 fluid simulations with different obstacle configurations were generated, each with 30 time steps (**Figure 1**), split into 900 training and 100 testing simulations using PhiFlow's pixel-based Navier-Stokes fluid modeling. Each simulation used a 12 by 24 pixel domain with 2x2 obstacles (**Figure 3**). To test how generalizable the network was, different sets of randomly located obstacles were used for each of the 1000 simulations.

In every simulation, each one of the 30 frames was paired with the next one, totaling 26,100 data inputs (with 2,900 data inputs for testing). The neural networks were implemented using PyTorch and take the first frame in each pair as input and predicts the *difference* between the two consecutive paired frames. The neural networks were trained using a mean-square error loss, for a total of over 2,000 training epochs with randomized batches, each containing 300 images. Multiple depths (i.e., number of layers) and widths (i.e., number of nodes per layer) were tested. To make predictions over *n > 1* time-steps, the neural networks are called *n* times, each time predicting the next frame (or more precisely the increment with respect to the previous frame) based on the previously predicted frame.

In all simulations, fluid "pours" up from a row at the bottom of the spatial domain with a set velocity that changes as it encounters obstacles. Fluid then exits through the "top" of the domain. The fluid entering the simulation has no difference in properties from the fluid originally inside the simulation at time $t$ = 0. Fluid cannot flow out through the side "walls" of the domain, creating areas with high velocity between the obstacles and "walls". Velocity is represented as a staggered grid in PhiFlow while the input fluid is represented by a centered grid. Note that divergence is present in the fluid because of the uneven acting forces from obstacles within the simulation.

**REFERENCES**

1. McCracken, Megan F. "Artificial neural networks in fluid dynamics: A novel approach to the Navier-Stokes equations." *Proceedings of the Practice and Experience on Advanced Research Computing: Seamless Creativity*. 2018. 1-4.
2. Cai, Kang, and Jiayao Wang. "Physics-Informed Neural Networks for Solving Incompressible Navier–Stokes Equations in Wind Engineering." *Physics of Fluids*, vol. 36, no. 12, 2024, article 121303, AIP Publishing. DOI: 10.1063/5.0244094
3. Zhu, Yongzheng, et al. "Physics-Informed Neural Networks for Unsteady Incompressible Flows with Time-Dependent Moving Boundaries." *arXiv*, 25 Aug. 2023, arxiv.org/abs/2308.13219.
4. Kochkov, Dmitry, et al. *"Machine Learning–Accelerated Computational Fluid Dynamics." Proceedings of the National Academy of Sciences*, vol. 118, no. 15, 2021, p. e2101784118.
5. Chen, Yu, et al. "A Pioneering Neural Network Method for Efficient and Robust Fluid Simulation." *arXiv*, 12 Dec. 2024, arxiv.org/abs/2412.10748.
6. Vinuesa, Ricardo, et al. "The Potential of Machine Learning to Enhance Fluid Mechanics." *Nature Reviews Physics*, vol. 3, no. 8, 2021, pp. 466–480. https://doi.org/10.1038/s41586-021-00344-2.

**Figures, Tables, and Captions**

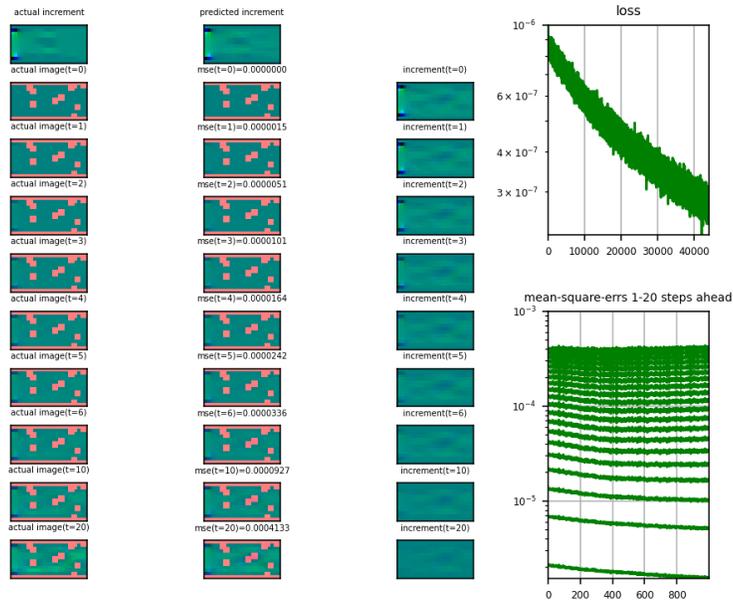

**Figure 1. Visualization of Navier-Stokes-simulated and neural network-simulated velocity graphs.** Column 1 illustrates fluid flow frames obtained through a numerical simulation of the Navier-Stokes equation. Column 2 shows the same frames as predicted by the neural network, with a mean square error measure displayed above each frame. Column 3 shows the increment (change) between each frame and the next one. Column 4 shows the evolution of the neural network loss function (mean square error in the y-axis of the top plot) and the mean-square error for 1 through 20 steps-ahead predictions (y-axis of the bottom plot) during training vs. the number training samples (x-axis).

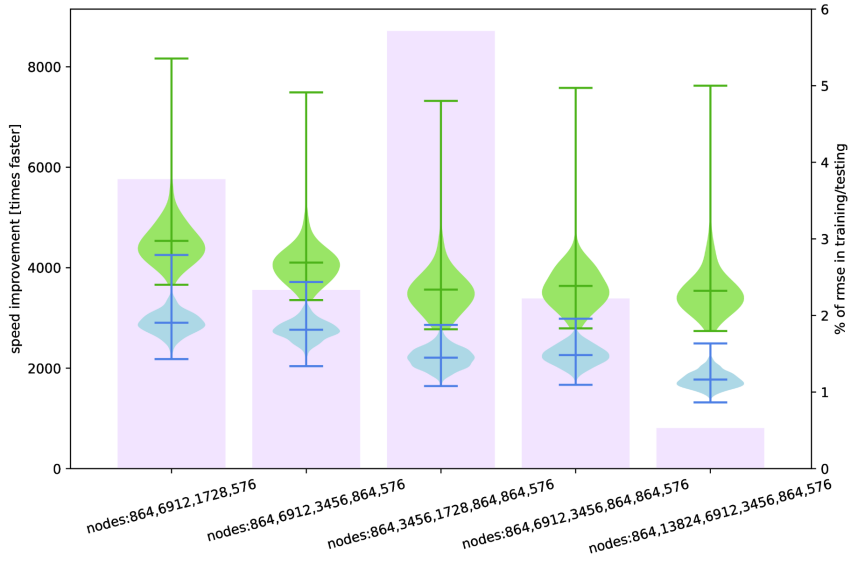

**Figure 2. Accuracy and efficiency of different neural networks.** The x-axis displays the number of nodes on each layer of 5 distinct neural networks . The left y-axis, corresponding to the pink bars, displays how many times the neural network is faster than a numerical simulation of the Navier-Stokes equation. The right y-axis, corresponding to the colored violin plots, displays the distribution of the root mean square error (RMSE), in percentage, for the different neural networks. The blue violin plots show the RMSE for the training data, while the green violin plots show the RMSE for the testing data.

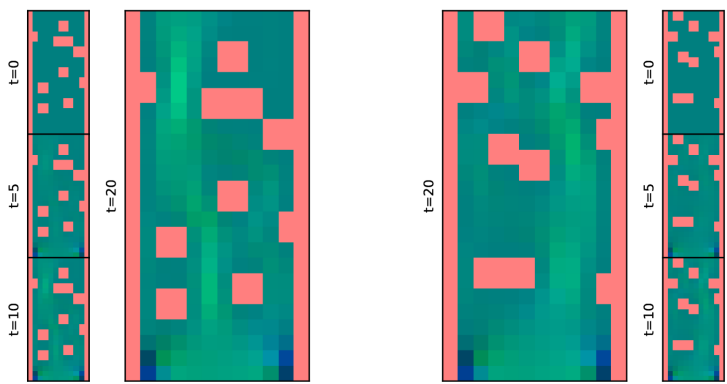

**Figure 3. Two PhiFlow simulations (one at left and one at right) showing a velocity graph and obstacles.** Red indicates an obstacle (where fluid cannot enter) and the red-blue shading the fluid velocity. Intense blue indicates a high velocity towards the left, and intense shades of green indicate a high velocity towards the upward direction. The velocity graphs are shown at different points in time, with the labels to the left of each figure showing the time frame *t*. The left simulation was taken from the training data while the right simulation was taken from the testing data set.

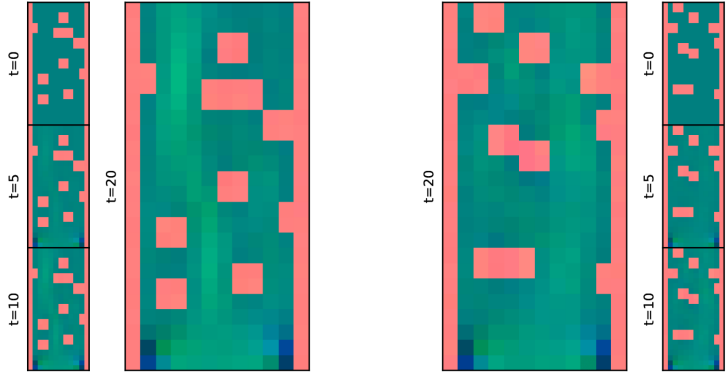

**Figure 4. Two neural network simulations (one at left and one at right) showing a velocity graph and obstacles.** The red color corresponds to obstacles (where fluid cannot enter) and the red-blue shading the fluid velocity. Intense blue indicates a high velocity towards the left, and intense shades of green indicate a high velocity upward. The velocity graphs are shown at different points in time, with the labels to the left of each figure showing the time frame *t*. Each frame in these simulation was generated sequentially by the neural network. The left simulation uses obstacles present in the training data while the right simulation uses obstacles from the testing data.

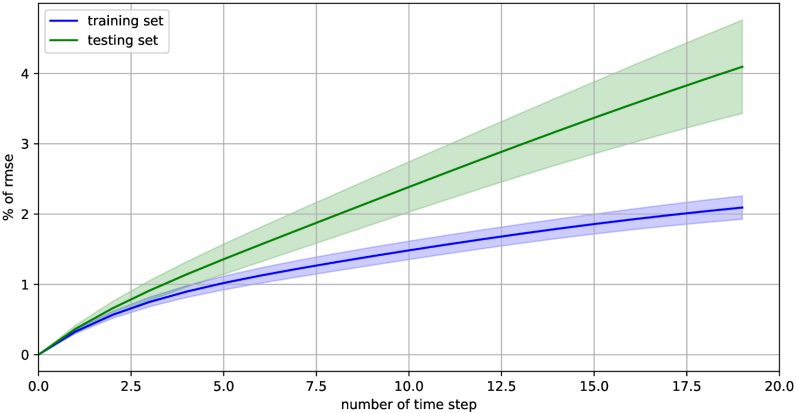

**Figure 5. Graph of (percentage) root mean square error in the neural network.** Line graph showing mean (solid lines) with standard deviations (shaded area) of the RMSE (in percentage,

in the y-axis) for the obstacles in the training (blue) and testing (green) data sets vs. the number of time steps over which the neural network is doing the prediction ( in the x-axis)

**Appendix**

https://github.com/ruihespanha/new-physics.git